\documentclass[journal]{IEEEtran}
\usepackage{amsmath,amsfonts}
\usepackage{algorithm}
\usepackage{algorithmic}

\usepackage{array}
\usepackage[caption=false,font=normalsize,labelfont=sf,textfont=sf]{subfig}
\usepackage{textcomp}
\usepackage{stfloats}
\usepackage{url}
\usepackage{verbatim}
\usepackage{graphicx}
\graphicspath{{figures/}}
\usepackage{multirow}
\usepackage{booktabs}
\usepackage[table,xcdraw]{xcolor}
\usepackage{float} 
\usepackage{cite}
\usepackage{lineno,hyperref}

\def\BibTeX{{\rm B\kern-.05em{\sc i\kern-.025em b}\kern-.08em
    T\kern-.1667em\lower.7ex\hbox{E}\kern-.125emX}}
\usepackage{balance}

\colorlet{myorange}{orange!85!black}
\newif\ifshowDL
\showDLtrue  

\begin{document}
\title{Towards Imperceptible Adversarial Defense: A Gradient-Driven Shield against Facial Manipulations}
\author{Yue Li, Linying Xue, Dongdong Lin, Qiushi Li, Hui Tian$^{\ast}$, \IEEEmembership{Senior Member, IEEE}, Hongxia Wang, \IEEEmembership{Member, IEEE}
\thanks{ 
Yue Li, Linying Xue, Dongdong Lin and Hui Tian are with the College of Computer Science and Technology, National Huaqiao University, Xiamen 361021, China, and also with the Xiamen Key Laboratory of Data Security and Blockchain Technology, Xiamen 361021, China (e-mail:  liyue\_0119@hqu.edu.cn; 23014083061@stu.hqu.edu.cn; dongdonglin8@gmail.com; htian@hqu.edu.cn).}
\thanks{Qiushi Li is with Media Integration and Communication Center (MICC), University of Florence, Florence, Italy. (email:
qiushi.li@unifi.it)}
\thanks{Hongxia Wang is with with the School of Cyber Science and Engineering,
Sichuan University, Chengdu 610207, China (e-mail:hxwang@scu.edu.cn)}
\thanks{\textit{Corresponding author: Hui Tian}}
}

\markboth{Journal of \LaTeX\ Class Files,~Vol.~xx, No.~x, xx~20xx}%
{How to Use the IEEEtran \LaTeX \ Templates}

\maketitle

\begin{abstract}
With the flourishing prosperity of generative models, manipulated facial images have become increasingly accessible, raising concerns regarding privacy infringement and societal trust. In response, proactive defense strategies embed adversarial perturbations into facial images to counter deepfake manipulation. However, existing methods often face a trade-off between imperceptibility and defense effectiveness—strong perturbations may disrupt forgeries but degrade visual fidelity. Recent studies have attempted to address this issue by introducing additional visual loss constraints, yet often overlook the underlying gradient conflicts among losses, ultimately weakening defense performance.
To bridge the gap, we propose a \underline{gr}adient-projection-based \underline{a}dver\underline{s}arial \underline{p}roactive defense (GRASP) method that effectively counters facial deepfakes while  minimizing perceptual degradation. 
GRASP is the first approach to successfully integrate both structural similarity loss and low-frequency loss to enhance perturbation imperceptibility. By analyzing gradient conflicts between defense effectiveness loss and visual quality losses, GRASP pioneers the design of the gradient-projection mechanism to mitigate these conflicts, enabling balanced optimization that preserves image fidelity without sacrificing defensive performance. Extensive experiments validate the efficacy of GRASP, achieving a PSNR exceeding 40 dB, SSIM of 0.99, and a 100\% defense success rate against facial attribute manipulations, significantly outperforming existing approaches in visual quality.
\end{abstract}


\begin{figure}[!t]
\centering
\includegraphics[width=\linewidth]{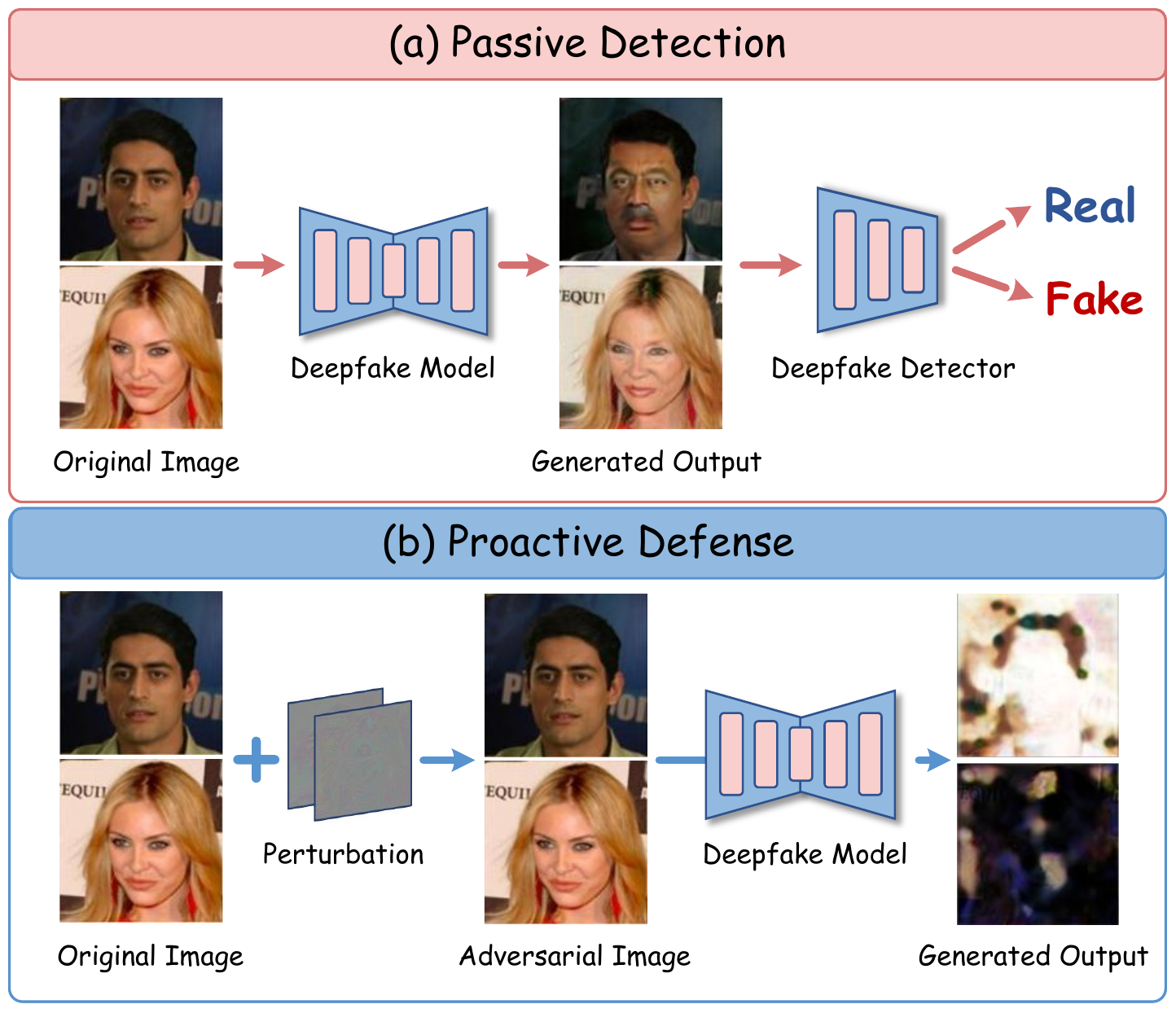}
    \caption{Diagrams of passive detection and proactive defense. (a) Passive Detection: A detector is employed to determine whether an image is forged. (b) Proactive Defense: Perturbations are added to the original image to disrupt the forgery process of deepfake models.}
\label{fig1}
\end{figure}

\section{Introduction}
\IEEEPARstart{W}{ith} the rapid development of deep learning technologies---particularly the emergence of generative models such as Generative Adversarial Networks (GANs)\cite{NIPS2014_5ca3e9b1}---the creation of images and videos has undergone a profound transformation. One prominent application is \textit{deepfake} technology\cite{wang2024deepfake}, which manipulates facial images to generate realistic faces with altered poses, emotions, expressions, or gender attributes. 
While technically impressive, the misuse of deepfakes has raised pressing concerns, as manipulated media involving celebrity figures can spread rapidly, leading to emotional distress and undermining trust in critical domains such as politics, law, and journalism \cite{10.1007/s11263-022-01606-8}. These risks underscore the urgent need for effective countermeasures to preempt potential misuse and safeguard both individual privacy and public trust. 

As a response to the deepfake threat, researchers have proposed two main defense strategies: \textit{passive detection} and \textit{proactive defense}, as illustrated in Fig. \ref{fig1}. Passive detection methods \cite{Wang_2020_CVPR,10.1145/3442381.3449809,10008209,10168141,11098842,10908403} primarily rely on well-trained deepfake detectors to analyze forged features of manipulated media, as shown in Fig. \ref{fig1}(a). While detection accuracy has steadily improved, this strategy presents notable limitations. First, as a post-hoc measure, it cannot prevent the spread of forged content or mitigate its impact during propagation. 
Second, passive detection may inadvertently drive the advancement of deepfake technology by exposing detection blind spots~\cite{9893860}. Thus, passive detection alone is insufficient to fundamentally curb the misuse of deepfake technology.

Proactive defenses based on adversarial perturbations have been proposed as a supplementary approach to counter deepfake by disrupting the generation of malicious content before it is distributed \cite{10.1007/978-3-030-66823-5_14,Huang_Zhang_Zhou_Zhang_Yu_2021,10100731,10086559,Huang2021CMUAWatermarkAC,ijcai2022p107,10.1145/3653457,qiao2024scalable,10008794,10121622,10330601,9710148,10.1609/aaai.v37i12.26693,10458678}, as illustrated in Fig. \ref{fig1}(b). The key challenge for proactive defense methods lies in achieving a balance between defense effectiveness and the imperceptibility of perturbations, as first articulated by Huang \textit{et al. } \cite{Huang_Zhang_Zhou_Zhang_Yu_2021}. Imperceptibility requires that the adversarial facial image remains visually indistinguishable from the original. In contrast, defense effectiveness demands that the introduced perturbations significantly impair the ability of deepfake models to synthesize realistic forgeries.


Existing methods have demonstrated promising defense effectiveness. Ruiz \textit{et al. } \cite{10.1007/978-3-030-66823-5_14} is the first to leverage adversarial perturbations for proactive defense against facial deepfakes. To further enhance defense effectiveness across multiple models, Huang \textit{et al.} \cite{Huang2021CMUAWatermarkAC} generate adversarial perturbations from each individual model and fuse them into a universal perturbation. Similarly, Tang \textit{et al.} \cite{10.1145/3653457} develop a gradient-ensemble strategy to enhance the overall perturbation impact. Although these approaches enhance cross-model defense performance, they often result in a noticeable degradation of visual quality in the adversarial images. 

For better visual quality, Li \textit{et al.} \cite{10121622} constrained the region where perturbations are applied. Building on this idea, Zhang \textit{et al. }~\cite{10330601} introduced a fine-grained module to more precisely control the distribution and intensity of the perturbations. 
However, the aforementioned methods rely solely on Iterative Fast Gradient Sign Method (IFGSM) to determine the direction of gradient updates when generating perturbations, without considering that the choice of direction may impact both visual quality and defense effectiveness. Hence, achieving high visual quality in adversarial images while maintaining strong defense effectiveness remains a challenging and unresolved problem.

To this end, we propose GRASP (\underline{GR}adient-projection-based \underline{A}dver\underline{S}arial \underline{P}roactive defense), a novel method that generates perturbations capable of effectively hindering deepfake manipulations while minimizing visual distortion in the adversarial images. Specifically,  GRASP leverages structural similarity loss and low-frequency loss to maintain visual quality, while mean squared error (MSE) loss is employed to achieve defense effectiveness. However, the simultaneous optimization of these objectives introduces inevitable gradient conflicts—visual quality losses encourage similarity to the original image, whereas the defense loss promotes distinction.
To resolve this, GRASP designs a gradient projection strategy based on normal vectors, which projects each gradient onto the normal plane of the others, yielding a conflict-free subspace for perturbation updates. Additionally, a Gaussian filtering layer is integrated into the perturbation generation process to further improve robustness. The main contributions are as follows.

\begin{itemize}
\item \textbf{A unified adversarial defense method GRASP} is proposed that achieves a high defense success rate while significantly improving both subjective and objective visual quality of adversarial facial images.

\item \textbf{A dual-perspective visual fidelity preservation mechanism} is introduced by combining structural similarity and low-frequency constraints, effectively retaining facial texture and minimizing visual degradation.

\item \textbf{A novel conflict-free gradient projection strategy} is designed to resolve inconsistent gradient directions among loss functions, enabling effective perturbation updates without introducing artifacts or noise.    
\end{itemize}


\section{Related Work}
\label{Related Work}
\subsection{Deep Facial Forgery}

Generative models have made remarkable advances in image synthesis, with GANs playing a particularly significant role in the development of deepfake technologies. By sampling from a random latent vector, models such as PGGAN~\cite{karras2018progressive} and  StyleGAN~\cite{karras2019style, karras2020analyzing} are capable of generating highly realistic yet non-existent facial images. In particular, StyleGAN supports fine-grained control over facial attributes, allowing the generation of forged faces with specific characteristics. 

Several advanced generative models have been developed to further enhance the controllability and realism of facial synthesis. StarGAN~\cite{8579014} takes both the input image and a domain label during training, incorporating a mask vector into the domain label to facilitate cross-domain style transfer. AttGAN~\cite{8718508} introduces an attribute classification constraint to improve the accuracy of attribute manipulations.  HiSD~\cite{9577868} offers an image-to-image translation framework that enables multi-label scalability and controllable diversity through unsupervised disentanglement of semantic attributes. Additionally, SimSwap~\cite{10.1145/3394171.3413630} exemplifies identity-level manipulation by transferring the facial identity from a source image to a target image while preserving contextual features such as expression and head pose.

\subsection{Proactive Defense Against Deepfake}

To mitigate the threats posed by deepfake technologies, researchers have proposed the concept of \emph{proactive defense}, which seeks to impede forgery at the content generation stage. 
One of the earliest methods in this field~\cite{10.1007/978-3-030-66823-5_14} involves embedding adversarial perturbations into facial images to interfere with the generation process, thereby substantially degrading the visual realism of the generated forgeries. Expanding on this idea, Huang \textit{et al.}~\cite{Huang_Zhang_Zhou_Zhang_Yu_2021} propose a framework that modifies facial data with imperceptible distortions, using a surrogate model to simulate the target deepfake system. These representative methods underscore the central challenge in proactive defense: achieving an effective trade-off between \emph{defense effectiveness} and \emph{perturbation imperceptibility}.

Initially, researchers put much effort into improving \emph{defense effectiveness}. 
Model-agnostic perturbation methods, such as CMUA~\cite{Huang2021CMUAWatermarkAC}, are proposed to maintain defense effectiveness across diverse forgery models. Lin \textit{et al.}~\cite{10008794} further explore the impact of perturbation injection order and introduce joint optimization strategies to enhance cross-model performance. Yeh \textit{et al.}~\cite{9710148} address the challenge of defending against unknown forgery models by proposing a perception-constrained, randomness-free gradient estimation approach, enabling the generation of adversarial perturbations without access to model gradients.
Ruiz \textit{et al.}~\cite{10.1609/aaai.v37i12.26693} significantly reduce the number of required queries by dynamically reusing previously generated perturbations. 
To avoid querying deepfake models, Dong \textit{et al.}~\cite{10100731} construct a substitute model based on face reconstruction, enabling the transfer of adversarial perturbations from the substitute to the inaccessible target models.
While the introduction of strong perturbation in these methods enhances defense effectiveness, it often comes at the cost of degraded visual quality.

To ensure \emph{perturbation imperceptibility}, several methods constrain perturbations to semantically meaningful facial regions. 
Li \textit{et al.}\cite{10121622} introduced saliency-aware mask to restrict perturbations to important facial region, improving perceptual quality. 
Zhang \textit{et al.}~\cite{10330601} used a union mask, combining a saliency mask and a manipulation mask, 
to guide perturbations toward critical facial regions. 
Qu \textit{et al.}~\cite{10458678} propose a robust adversarial perturbation method that maintain imperceptibility while resisting compression artifacts introduced by online social networks.


While prior adversarial defense methods have demonstrated either strong defense effectiveness or acceptable imperceptibility, they often struggle to achieve a satisfactory balance between visual quality and defense success rate, frequently sacrificing one to improve the other. The proposed GRASP aims to address this limitation by pursuing a defense strategy that simultaneously ensures high perceptual fidelity and strong resistance to deepfake manipulation.

\begin{figure*}[!t]
\centering
\includegraphics[width=\linewidth]{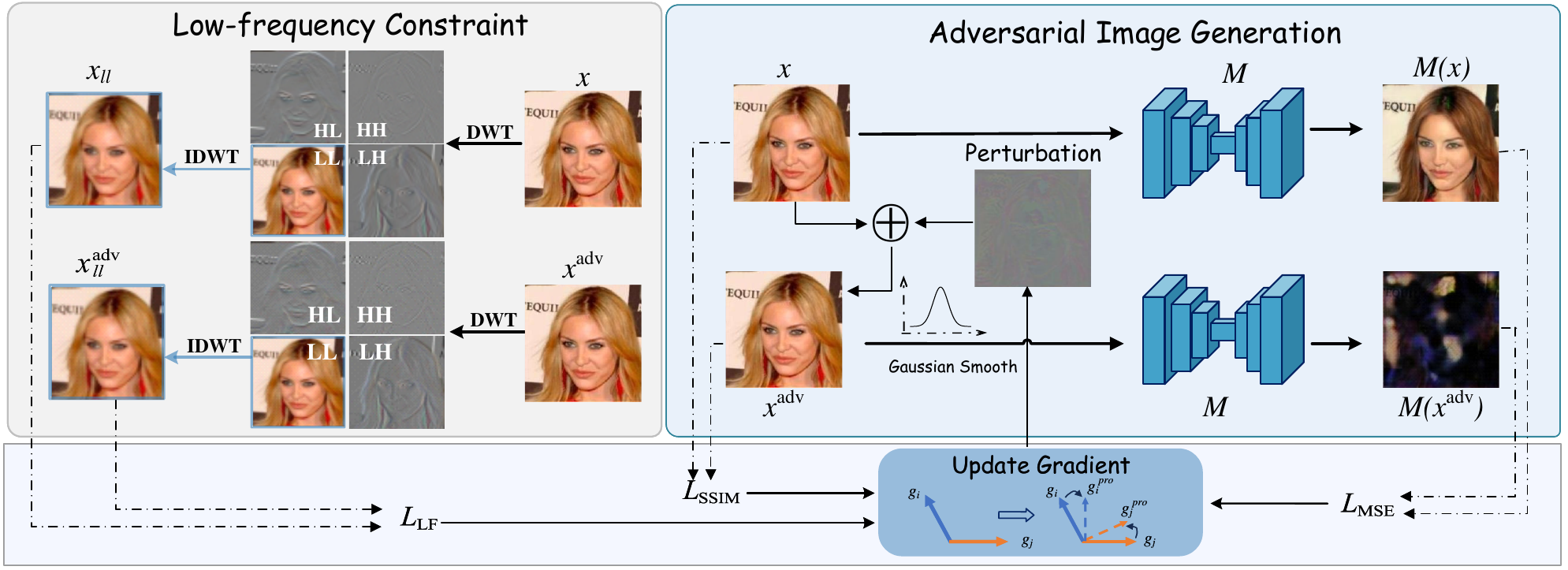}
\caption{Overview of GRASP: The proposed method enhances the MSE loss between the outputs of the forgery model when given original and adversarial facial images as input, while simultaneously minimizing the SSIM loss and low-frequency loss between the original and adversarial images. Gradient projection is employed to migrate gradient conflicts. \textbf{Low-Frequency Constraint} denotes the construction of the low-frequency loss, while \textbf{Adversarial Image Generation} illustrates the process of crafting adversarial facial images.
}
\label{fig:fig3}
\end{figure*}

\section{Problem Formulation}
\label{Problem Formulation}

Deepfake generation involves modifying or synthesizing facial content to produce manipulated outputs using a pre-trained generative model \( M \). Given an input image \( x \in X \), the model generates a manipulated face image \( y = M(x) \), where \( y \in Y \) exhibits semantic changes such as attribute editing or face swapping. Here, \(x \in X\) is a natural face image, and \(y \in Y\) is the corresponding manipulated output.

In this context, we consider a proactive defense scenario where the defender seeks to inject an imperceptible perturbation \(\eta\) into the original image \(x\), such that the resulting adversarial image \(x^{\text{adv}} = x + \eta\) degrades the output quality of the deepfake model \(M\). The goal is to render the manipulated image \(M(x^{\text{adv}})\) less realistic or semantically inconsistent, thereby undermining the effectiveness of deepfake generation. This task is subject to two requirements (RQs):
\begin{itemize}
    \item RQ1: \emph{The defense should remain effective across various deepfake models and attributes}.
    \item RQ2: \emph{The perturbation should be visually imperceptible}. 
\end{itemize}
To satisfy RQ1, the objective can be formulated as:
\begin{equation}
\max_{\|\eta\|_{\infty} \leq \epsilon} L(M(x), M(x+\eta)),
\label{eq:loss_function}
\end{equation}
where \(L(\cdot)\) denotes a distance metric (e.g., MSE) to quantify the degradation of the manipulated output. To satisfy RQ2, one has to ensure that the perturbation \(\eta\) is small enough, i.e., \(\|\eta\|_\infty \leq \epsilon\), where \(\epsilon\) is a small constant that limits the perturbation magnitude.

\section{Method}
\label{method}

In this section, the proposed proactive deepfake defense method, GRASP, is presented. We begin with an overview of the overall framework, followed by a detailed description of the loss function design. We then analyze the challenges posed by gradient conflicts among multiple loss function terms and introduce a gradient projection strategy to resolve this issue.

\subsection{Overview}
As shown in Fig.~\ref{fig:fig3}, the proposed GRASP framework formulates the generation of adversarial examples as a min-max optimization problem to satisfy the two RQs.

\emph{Defense effectiveness} (RQ1): The objective is to maximize the discrepancy between the original output and the disrupted output. This is achieved through an MSE loss between the output image generated by the deepfake model when fed with the original image $x$ and the adversarial image $x^{\text{adv}}$, respectively.

\emph{Perturbation imperceptibility} (RQ2): To minimize visual artifacts in the adversarial image, our method incorporates the SSIM loss and the low-frequency constraint on the original image $x$ and the adversarial image $x^{\text{adv}}$.

Simultaneous optimization of these losses leads to gradient conflicts, as each objective may induce competing update directions. To mitigate this, GRASP tailors a gradient projection strategy that resolves such conflicts by mutually projecting gradients onto each other's normal planes. This preserves both defense effectiveness and imperceptibility in the presence of conflicting gradients.

\subsection{Loss Function Design}
\label{Loss function design}

\emph{Defense effectiveness} (RQ1): The goal of the defense is to maximize the discrepancy between the original and manipulated outputs produced by the deepfake model. To evaluate how effectively the adversarial perturbation disrupts the manipulated output, MSE loss is calculated to measure the difference of these two outputs:
\begin{equation}
    L_\text{MSE}(x, x^{\text{adv}}) = \|M(x) - M(x^{\text{adv}})\|_2^2,
    \label{eq:mse_loss}
\end{equation}
where \( x \) represents the original facial image, and \( x^{\text{adv}} \) is the adversarial version with perturbations.

\emph{Perturbation imperceptibility} (RQ2): To mitigate the distortion of facial images caused by perturbations, we introduce a structural similarity loss, which is widely used in image processing to quantify the perceptual similarity between two images. It can be expressed as:
\begin{equation}
    L_{\text{SSIM}}(x, x^{\text{adv}}) = \frac{(2\mu_x \mu_{x^{\text{adv}}} + C_1)(2\sigma_{x} \sigma_{x^{\text{adv}}} + C_2)}{(\mu_x^2 + \mu_{x^{\text{adv}}}^2 + C_1)(\sigma_x^2 + \sigma_{x^{\text{adv}}}^2 + C_2)},
    \label{eq:ssim_loss}
\end{equation}
where \(\mu \) and \(\sigma\) represent the mean and variance, while \(C_1\) and \(C_2\) are small constants that prevent division by zero. 

To further improve the perceptual quality of adversarial images, we incorporate a low-frequency loss when updating the adversarial image. 
Since low-frequency components are more noticeable to the human visual system, suppressing perturbations in this frequency band helps reduce visual artifacts.
As shown in Fig.~\ref{fig:fig3}, Discrete Wavelet Transform (DWT) is applied to decompose the facial image \(x\) into four subbands, each corresponding to a different frequency component:
\begin{equation}
    \text{DWT}(x) = x^{ll},x^{lh},x^{hl},x^{hh},
\end{equation}
where the low-frequency component \(x_{ll}\) contains the main information of the facial image, while \(x^{lh}\), \(x^{hl}\), \(x^{hh}\) contain high-frequency details.
The low-frequency component \(x^{ll}\) is then used to reconstruct the image, which can be expressed as:
\begin{equation}\label{eq:lf}
    \phi(x) = \text{IDWT}(x^{ll}),
\end{equation}
where $ \text{IDWT}(\cdot)$ denotes the inverse discrete wavelet transform. The low-frequency loss~\cite{9879877} is then expressed as:
\begin{equation}
    L_{\text{LF}}(x, x^{\text{adv}}) = \|\phi(x) - \phi(x^{\text{adv}})\|_1.
    \label{eq:lf_loss}
\end{equation}

Finally, RQ1 and RQ2 are simultaneously solved by minimizing the following loss:
\begin{equation}
\begin{split}
    L = &-  L_{\text{MSE}}(M(x),M(x^{\text{adv}}))  \\
    &+   L_{\text{SSIM}}(x, x^{\text{adv}}) +  L_{\text{LF}}(x, x^{\text{adv}}).
    \label{eq:total_loss}
\end{split}
\end{equation}


\subsection{Defense Retained Adversarial Image Generation}
\label{Gradient Projection}

\begin{figure}[!t]
\centering
\includegraphics[width=\linewidth]{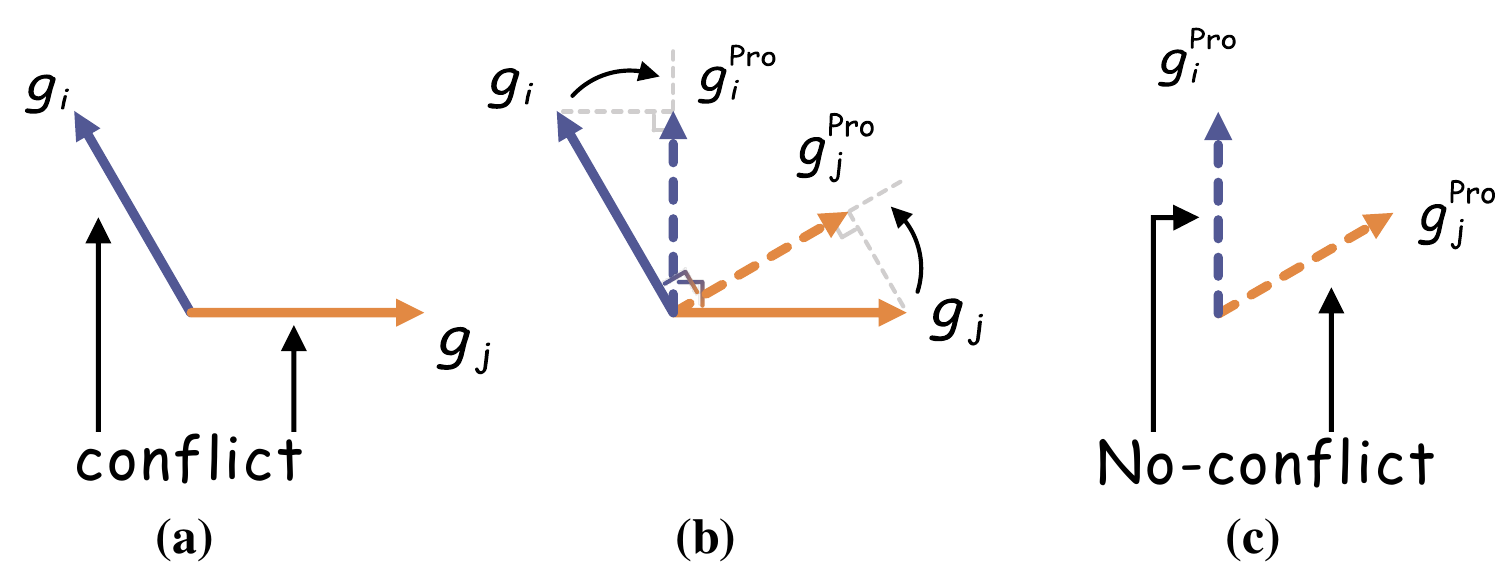}
\caption{Gradient projection strategy to resolve gradient conflicts. (a) The gradient directions \( g_i \) and \( g_j \) are in conflict. (b) To resolve the conflict, \( g_i \) and \( g_j \) are mutually projected onto each other's normal planes. (c) Conflict-free gradients after projection are indicated.}
\label{fig4}
\end{figure}

As presented in Section \ref{Loss function design}, the loss term $L_{\text{MSE}}$ (Eq.~\eqref{eq:mse_loss}) is designed to encourage output discrepancy, while $L_{\text{SSIM}}$ (Eq.~\eqref{eq:ssim_loss}) and $L_{\text{LF}}$ (Eq.~\eqref{eq:lf_loss}) are intended to preserve visual similarity. Although all three losses pertain to visual perception, they pursue inherently conflicting objectives. When jointly optimized as defined in Eq.~\eqref{eq:total_loss}, their gradient directions may conflict (see Fig.~\ref{fig4}(a)), potentially hindering effective convergence. Furthermore, such conflicts can adversely affect the defense performance of the resulting adversarial image, reducing its ability to effectively disrupt deepfake synthesis while maintaining visual quality.
Therefore, a key question arises: \emph{How can defense performance be retained when gradient conflicts occur?}

As the answer, the gradient projection strategy~\cite{NEURIPS2020_3fe78a8a, 10.1016/j.asoc.2023.111088} steps into the spotlight.  
PCGrad~\cite{NEURIPS2020_3fe78a8a} introduces a unidirectional gradient projection strategy in multi-task learning. NPGA~\cite{10.1016/j.asoc.2023.111088} extends this idea to adversarial attacks in classification models, applying PCGrad to emphasize imperceptibility. In each iteration, NPGA prioritizes projecting the perturbation gradient toward a direction that enhances visual stealth, effectively focusing on a single objective. 

To achieve the ultimate goal of GRASP—meeting both RQ1 and RQ2—we draw inspiration from PCGrad and NPGA and tailor the gradient projection strategy for adversarial image generation. Specifically, we adopt a cross-projection strategy (see Fig.~\ref{fig4}(b)) that simultaneously accounts for gradient interactions between loss terms. Unlike prior approaches that focus on a single objective, our method dynamically balances these competing goals during each iteration, ensuring that no single objective dominates the optimization process. 
As a result, the final projected gradient is effectively conflict-free, as illustrated in Fig.~\ref{fig4}(c).

The procedure of the adversarial image generation is presented in Algorithm~\ref{algorithm}. The three losses—$L_{\text{MSE}}$, $L_{\text{SSIM}}$, and $L_{\text{LF}}$—are jointly optimized to generate adversarial images.  Before assessing gradient conflicts, the gradients of the three losses are computed and normalized using the \( \ell_1 \)-norm for stability:
\begin{equation}
    g_t = -\frac{\nabla_{x_t} L_{\text{MSE}}(x, x_t^{\text{adv}})}{\|\nabla_{x_t} L_{\text{MSE}}(x, x_t^{\text{adv}})\|_1 + \xi}, \label{eq:g_t} \end{equation}
\begin{equation}    h_t = \frac{\nabla_{x_t} L_{\text{SSIM}}(x, x_t^{\text{adv}})}{\|\nabla_{x_t} L_{\text{SSIM}}(x, x_t^{\text{adv}})\|_1 + \xi}, \label{eq:h_t} \end{equation}
\begin{equation}    z_t = \frac{\nabla_{x_t} L_{\text{LF}}(x, x_t^{\text{adv}})}{\|\nabla_{x_t} L_{\text{LF}}(x, x_t^{\text{adv}})\|_1 + \xi} \label{eq:z_t}\end{equation}
where \(\xi\) is a small constant to prevent division by zero.
To mitigate potential conflicts between two gradients \(\mathbf{a}\) and \(\mathbf{b}\), the tailored gradient projection strategy is applied only when their inner product is non-positive. The resulting gradient is then defined as:
\begin{equation}\label{eq:gradient_projection}
G(\mathbf{a}, \mathbf{b}; \lambda, \mu) =
\begin{cases}
    \lambda\, \text{Proj}_\mathbf{b}\mathbf{a} + \mu\, \text{Proj}_\mathbf{a}\mathbf{b} , & \text{if } \langle \mathbf{a}, \mathbf{b} \rangle \leq 0, \\
    \lambda\, \mathbf{a} + \mu\, \mathbf{b}, & \text{otherwise},
\end{cases}
\end{equation}
where the projection onto the normal plane is given by
\begin{equation}
    \text{Proj}_\mathbf{b}\mathbf{a} = \mathbf{a} - \frac{\langle \mathbf{a}, \mathbf{b} \rangle}{\|\mathbf{b}\|^2} \mathbf{b}, \quad 
    \text{Proj}_\mathbf{a}\mathbf{b} = \mathbf{b} - \frac{\langle \mathbf{b}, \mathbf{a} \rangle}{\|\mathbf{a}\|^2} \mathbf{a},
\end{equation}
\(\lambda\) and \(\mu\) are hyperparameters controlling the contribution of each gradient.

Then, by substituting the gradients \(g_t\), \(h_t\), and \(z_t\) into Eq.~\eqref{eq:gradient_projection}, three new projected gradients are obtained:
\begin{align}
    G_t^{s_1} &= G(g_t, h_t; \lambda_1, \mu_1), \label{eq:gs1}  \\
    G_t^{s_2} &= G(h_t, z_t; \lambda_2, \mu_2), \label{eq:gs2}  \\
    G_t^{s_3} &= G(g_t, z_t; \lambda_3, \mu_3), \label{eq:gs3} 
\end{align}
and the total conflict-free gradient is
\begin{equation}\label{eq:G_total}
    G_t^{\text{total}} = \eta_1\, G_t^{s_1} + \eta_2\, G_t^{s_2} + \eta_3\, G_t^{s_3},
\end{equation}
where \(\eta_1\), \(\eta_2\), and \(\eta_3\) are hyperparameters that control the strength of each projected gradient.

Finally, $G_t^{\text{total}}$ is used to update the adversarial image in $t$-th iteration, as follows:
\begin{equation}\label{eq:update_adv_image}
    x_{t+1}^{\text{adv}} = \text{Clip}_{\epsilon}(x_{t}^{\text{adv}} + \kappa G_t^{\text{total}}),
\end{equation}
where \(\text{Clip}_{\epsilon}(\cdot)\) denotes an element-wise clipping function that constrains the perturbation within \([-\epsilon, +\epsilon]\), and \(\kappa\) is a fixed hyperparameter. Gaussian smoothing \cite{10.1007/978-3-030-66823-5_14} is applied to the adversarial image $x_{t+1}^{\text{adv}}$ at each iteration to enhance robustness against image transformations.

\begin{algorithm}[htbp]
    \caption{Adversarial Image Generation}
    \renewcommand{\algorithmiccomment}[1]{$//$ #1}
    \label{algorithm}
    \textbf{Input:} Target model $M$, original image $x$, loss functions $L_{\text{MSE}}, L_{\text{SSIM}}, L_{\text{LF}}$, maximum perturbation $\epsilon$, number of iterations $T$, hyperparameters $\lambda_1$, $\mu_1$, $\lambda_2$, $\mu_2$, $\lambda_3$, $\mu_3$, $\eta_1$, $\eta_2$, $\eta_3$, $\kappa$.\\
    \textbf{Output:} Adversarial image $x^{\text{adv}}$
    \begin{algorithmic}[1]
        \STATE Initialize $x_0^{\text{adv}} \leftarrow x$
        \FOR{$t = 0, 1, \dots, T-1$}
            \STATE Compute gradients $g_t, h_t, z_t$ via Eqs.~(\ref{eq:g_t})--(\ref{eq:z_t})
            \STATE $G_t^{s_1} \leftarrow G(g_t, h_t; \lambda_1, \mu_1)$
            \STATE $G_t^{s_2} \leftarrow G(h_t, z_t; \lambda_2, \mu_2)$
            \STATE $G_t^{s_3} \leftarrow G(g_t, z_t; \lambda_3, \mu_3)$
            \STATE $G_t^{\text{total}} \leftarrow \eta_1 G_t^{s_1} + \eta_2 G_t^{s_2} + \eta_3 G_t^{s_3}$
            \STATE $x_{t+1}^{\text{adv}} \leftarrow \text{Clip}_\epsilon(x_t^{\text{adv}} + \kappa G_t^{\text{total}})$
            \STATE Apply Gaussian smoothing to $x_{t+1}^{\text{adv}}$
        \ENDFOR
        \STATE $x^{\text{adv}} \leftarrow x_T^{\text{adv}}$
        \RETURN $x^{\text{adv}}$
    \end{algorithmic}
\end{algorithm}

\section{Experiments}
\label{exre}

This section presents a series of experiments to demonstrate the effectiveness of the proposed method in achieving high defense performance while preserving visual quality. We begin by introducing the experimental setup in Section~\ref{Experimental Setup} to ensure clarity and reproducibility. The performance of the method across different data scales is then illustrated using quantitative charts.  Section~\ref{Performance Comparison} presents a comparative analysis of defense effectiveness, perturbation imperceptibility, and robustness across attribute editing and face swapping scenarios. Finally, an ablation study and parameters analysis in Section~\ref{Ablation Study} evaluate the contributions of the proposed gradient projection strategy and justify the chosen hyperparameter settings.

\subsection{Experimental Setup}
\label{Experimental Setup}

\subsubsection{Deepfake Models}

In this paper, we evaluate three facial attribute editing models—StarGAN~\cite{8579014}, AttGAN~\cite{8718508}, and HiSD~\cite{9577868}—as well as one face swapping model, SimSwap \cite{10.1145/3394171.3413630}. StarGAN and AttGAN are configured to manipulate five attributes: \{black hair, brown hair, blond hair, gender, age\}, while HiSD adopts five attributes: \{blond hair, black hair, brown hair, bangs, glasses\}. All models are evaluated using their official pretrained weights and experimental settings as provided in their respective original papers.


\begin{figure*}[!t]
    \centering
    \includegraphics[width=\textwidth]{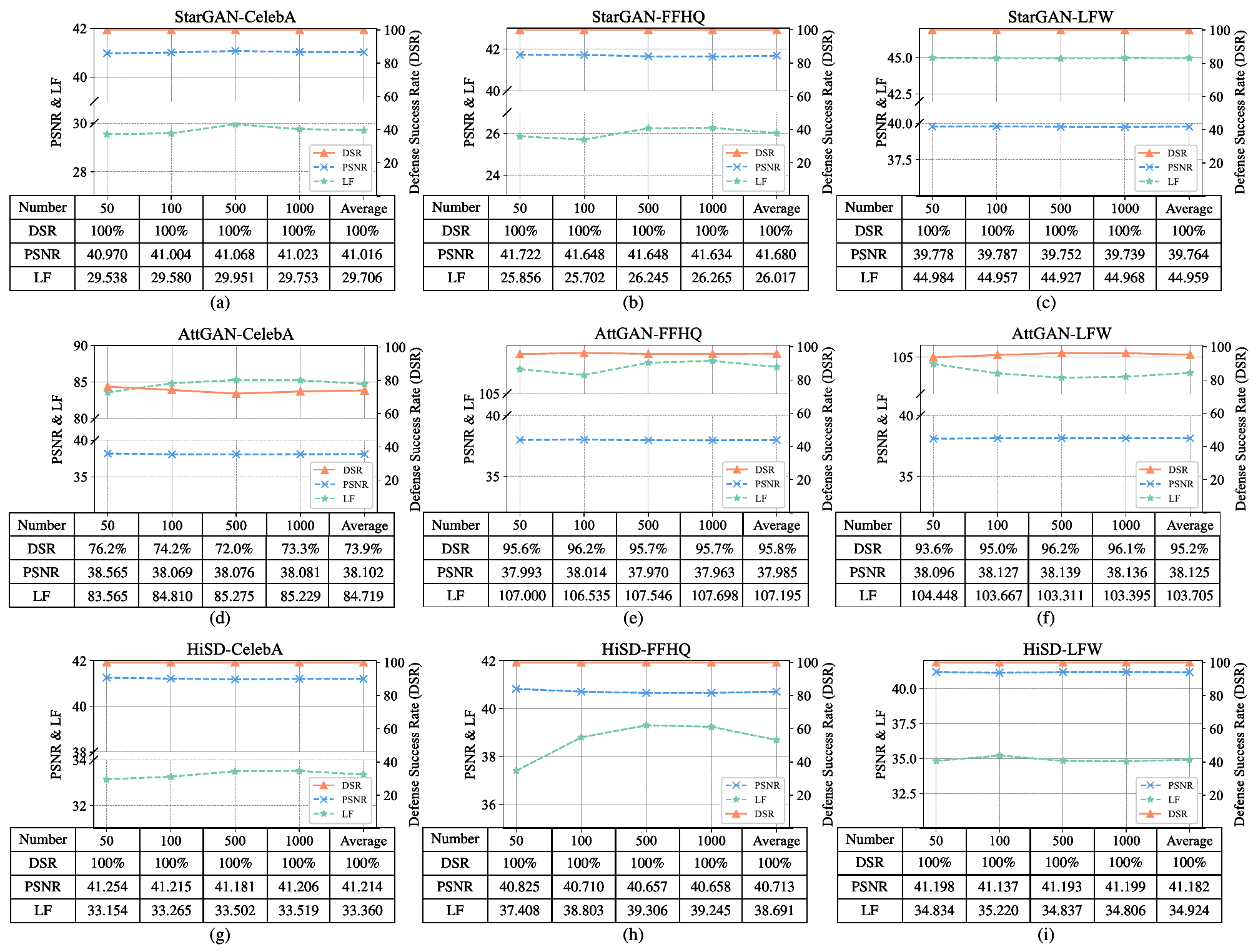}
    \captionsetup{justification=centering}
    \caption{The figure presents the experimental results of GRASP across different models and datasets. Subfigures (a)-(c) illustrate the performance of GRASP on the StarGAN model using the CelebA, FFHQ, and LFW datasets, evaluated under varying numbers of input images (50, 100, 500, 1000, and the overall average) with respect to the DSR, PSNR, and LF metrics. Subfigures (d)-(f) report the corresponding results for the AttGAN model under the same settings, while subfigures (g)-(i) present the outcomes for the HiSD model. }
    \label{three_images}
\end{figure*}

\subsubsection{Defense Methods for Comparison}

Five proactive deepfake defense methods are selected for comparison, including White-blur\cite{10.1007/978-3-030-66823-5_14}, AF\cite{ijcai2022p107}, Saliency-aware\cite{10121622}, Union-aware\cite{10330601} and DF-RAP\cite{10458678}. All of these methods aim to achieve a balance between perturbation imperceptibility and defense effectiveness. 
White-blur is the first method to leverage adversarial perturbation, using MSE loss as its primary optimization objective. 
Building on this, Saliency-aware restricts perturbations to salient facial regions to minimize visual artifacts. Union-aware adds a noise generation module and includes an SSIM loss to enhance image quality. AF generates perturbations in the Lab color domain to degrade the visual quality of the forged image. DF-RAP employs a pre-trained compression module to resist compression artifacts introduced by online social networks.

\subsubsection{Datasets}
The Datasets utilized for model training and adversarial image generation include CelebA \cite{10.1109/ICCV.2015.425}, FFHQ \cite{karras2019style}, and LFW \cite{huang2008labeled}. The CelebA dataset comprises over 200,000 celebrity images annotated with 40 facial attributes. FFHQ contains 70,000 high-quality images with exceptionally high resolution and diverse visual characteristics. LFW includes 13,233 real-world facial images, capturing a wide range of variations in pose, lighting, and expression. For defense evaluation, 100 images are randomly selected from each dataset as the testing set.


\subsubsection{Evaluation Metrics}
\label{Evaluation Metrics}

To evaluate the \emph{defense effectiveness (RQ1)} of the method, we adopt Defense Success Rate (DSR) and \(L_2\) loss as the primary metrics. Specifically, following the criterion established in \cite{10.1007/978-3-030-66823-5_14}, a defense is deemed successful if the \(L_2\) distance between the original and adversarial outputs of the deepfake model \(M\) exceeds 0.05. The DSR is defined as the proportion of adversarial images satisfying this condition, i.e.,
\begin{equation}
\text{DSR} = \frac{1}{N} \sum_{i=1}^{N} \mathbb{I}\left[\left\| M(x_i) - M(x_i^{\text{adv}}) \right\|_2 > 0.05\right],
\end{equation}
where \(x_i\) and \(x_i^{\text{adv}}\) denote the original and adversarial inputs respectively, \(N\) is the total number of adversarial images, and \(\mathbb{I}[\cdot]\) is the indicator function.

For the evaluation of \emph{perturbation imperceptibility (RQ2)}, four metrics to quantify the differences between original facial images \(x\) and adversarial images \(x^{\text{adv}}\) are utilized, they are: Peak Signal-to-Noise Ratio (PSNR), Structural Similarity Index (SSIM), Learned Perceptual Image Patch Similarity (LPIPS), and low-frequency distortion (LF). LPIPS computes perceptual similarity by measuring feature distances between images using pre-trained deep networks (e.g., VGG~\cite{simonyan2015deepconvolutionalnetworkslargescale}), closely aligning with human visual perception. LF quantifies the average low-frequency discrepancy:
\begin{equation}
\text{LF} = \frac{1}{N} \sum_{i=1}^{N} \left\| \phi(x_i) - \phi(x_i^{\text{adv}}) \right\|^2,
\end{equation}
where \(\phi(\cdot)\) is the low-frequency component (see Eq.~\eqref{eq:lf}).



\begin{table*}[htbp]
    \centering
    \tiny
    \renewcommand{\arraystretch}{1.1}
    \caption{Comparative results for defense methods across different datasets and deepfake models \\($\uparrow$ means the higher the better, $\downarrow$ means the lower the better)}
    \label{table1}
    \small
    \begin{tabular}{c|c|l|cc|cccc}
    \hline
    Model     & Datasets     & Method         & DSR↑                &  $L_2$↑       & PSNR↑         & SSIM↑        & LPIPS↓       & LF↓         \\ \hline
    &                        & White-blur\cite{10.1007/978-3-030-66823-5_14}     & \textbf{100\%}      & \underline{0.577}     & 34.198        & 0.937        & 0.033        & 119.292\\
    &                        & AF\cite{ijcai2022p107}     & \textbf{100\%}      & \textbf{0.968}    & \textbf{42.885}        & \textbf{0.990}        & \textbf{0.001}        & \textbf{4.494}   \\
    &                        & Saliency-aware\cite{10121622} & \textbf{100\%}      & 0.354     & 34.909        & 0.957        & 0.024        & 106.361                                 \\
    &                        & Union-aware\cite{10330601}    & 98.2\%              & 0.386     & 37.256        & 0.961        & 0.021        & 72.890                                  \\
    &                        & DF-RAP\cite{10458678}         & 99.6\%               & 0.265          & 34.993              & 0.945              & 0.104           &  73.936                            \\
    & \multirow{-5}{*}{CelebA\cite{10.1109/ICCV.2015.425}} & \cellcolor[HTML]{EFEFEF}GRASP & \cellcolor[HTML]{EFEFEF}\textbf{100\%} & \cellcolor[HTML]{EFEFEF}0.465 & \cellcolor[HTML]{EFEFEF}\underline{39.783} & \cellcolor[HTML]{EFEFEF} \underline{0.986} & \cellcolor[HTML]{EFEFEF}\underline{0.008} & \cellcolor[HTML]{EFEFEF}\underline{42.452} \\ \cline{2-9}
    &                        & White-blur\cite{10.1007/978-3-030-66823-5_14}     & \textbf{100\%}      & \underline{0.516}     & 34.212        & 0.943        & 0.035        & 122.494                                 \\
    &                        & AF\cite{ijcai2022p107}     & \textbf{100\%}      & \textbf{0.927}     & \textbf{44.185}        & \textbf{0.991}        & \textbf{0.001}        & \textbf{2.447}\\
    &                        & Saliency-aware\cite{10121622} & \textbf{100\%}      & 0.383     & 33.824        & 0.945        & 0.024        & 142.607                                 \\
    &                        & Union-aware\cite{10330601}    & 96.2\%              & 0.352     & 37.057        & 0.964        & 0.015        & 76.020                                  \\
    &                        & DF-RAP\cite{10458678}         & 100\%              & 0.357     & 33.954       & 0.936        & 0.180       &  67.485                            \\
    & \multirow{-5}{*}{FFHQ\cite{karras2019style}}   & \cellcolor[HTML]{EFEFEF}GRASP & \cellcolor[HTML]{EFEFEF}\textbf{100\%} & \cellcolor[HTML]{EFEFEF}0.419 & \cellcolor[HTML]{EFEFEF}\underline{39.413} & \cellcolor[HTML]{EFEFEF}\underline{0.986} & \cellcolor[HTML]{EFEFEF}\underline{0.007} & \cellcolor[HTML]{EFEFEF}\underline{47.167} \\ \cline{2-9} 
    &                        & White-blur\cite{10.1007/978-3-030-66823-5_14}     & \textbf{100\%}     & \underline{0.495}      & 34.506        & 0.937        & 0.035        & 117.671                                 \\
    &                        & AF\cite{ijcai2022p107}     & \textbf{100\%}      & \textbf{0.878}     & \textbf{43.256}        & \textbf{0.991}        & \textbf{0.001}        & \textbf{3.451}\\
    &                        & Saliency-aware\cite{10121622} & \textbf{100\%}     & 0.320      & 35.139        & 0.960        & 0.024        & 106.311                                 \\
    &                        & Union-aware\cite{10330601}    & 99.4\%             & 0.349      & 37.035        & 0.961        & 0.097        & 73.005                                  \\
    &                        & DF-RAP\cite{10458678}         & 100\%              & 0.262      & 34.908        & 0.946        & 0.146       &  69.274 \\
    \multirow{-15}{*}{StarGAN\cite{8579014}} & \multirow{-4}{*}{LFW\cite{huang2008labeled}}    & \cellcolor[HTML]{EFEFEF}GRASP & \cellcolor[HTML]{EFEFEF}\textbf{100\%} & \cellcolor[HTML]{EFEFEF}0.393 & \cellcolor[HTML]{EFEFEF}\underline{39.789} & \cellcolor[HTML]{EFEFEF}\underline{0.984} & \cellcolor[HTML]{EFEFEF}\underline{0.009} & \cellcolor[HTML]{EFEFEF}\underline{44.951} \\ \hline
    &                        & White-blur\cite{10.1007/978-3-030-66823-5_14}     & \underline{74.0\%}    & 0.121  & 33.597   & 0.936        & 0.163        & 265.672                                 \\
    &                        & AF\cite{ijcai2022p107}     & 14.5\%      & 0.024     & \underline{36.916}        & \underline{0.968}        & \textbf{0.044}        & \underline{111.601}\\
    &                        & Saliency-aware\cite{10121622} & 61.0\%             & 0.095           & 34.570   & 0.956        & 0.104        & 222.946                                 \\
    &                        & Union-aware\cite{10330601}    & 52.2\%             & 0.131           & 36.817   & 0.959        & 0.101        & 152.799                                 \\
    &                        & DF-RAP\cite{10458678}         & 71.8\%               & \underline{0.143}          & 34.652              & 0.944              & 0.174           &  137.339                            \\
    & \multirow{-5}{*}{CelebA\cite{10.1109/ICCV.2015.425}} & \cellcolor[HTML]{EFEFEF}GRASP & \cellcolor[HTML]{EFEFEF}\textbf{76.2\%}         & \cellcolor[HTML]{EFEFEF}\textbf{0.155}          & \cellcolor[HTML]{EFEFEF}\textbf{38.183} & \cellcolor[HTML]{EFEFEF}\textbf{0.978} & \cellcolor[HTML]{EFEFEF}\underline{0.060} & \cellcolor[HTML]{EFEFEF}\textbf{83.565} \\  \cline{2-9} 
    &                        & White-blur\cite{10.1007/978-3-030-66823-5_14}     & \underline{96.8\%}    & 0.189           & 33.647   & 0.943        & 0.116        & 259.786                                 \\
    &                        & AF\cite{ijcai2022p107}     & 55.2\%      & 0.107     & \underline{36.232}        & \underline{0.966}        & \textbf{0.027}        & \underline{115.651}\\
    &                        & Saliency-aware\cite{10121622} & \textbf{97.4\%}             & \underline{0.192}  & 33.118   & 0.943        & 0.106        & 301.799                                 \\
    &                        & Union-aware\cite{10330601}    & 72.9\%             & 0.183           & 35.418   & 0.955        & 0.083        & 183.147                                 \\
    &                        & DF-RAP\cite{10458678}         & 96.0\%               & \textbf{0.236}          & 34.597              & 0.951              & 0.112           &  133.515                            \\
    & \multirow{-5}{*}{FFHQ\cite{karras2019style}}   & \cellcolor[HTML]{EFEFEF}GRASP & \cellcolor[HTML]{EFEFEF}96.2\%         & \cellcolor[HTML]{EFEFEF}0.167          & \cellcolor[HTML]{EFEFEF}\textbf{38.014} & \cellcolor[HTML]{EFEFEF}\textbf{0.985} & \cellcolor[HTML]{EFEFEF}\underline{0.044} & \cellcolor[HTML]{EFEFEF}\textbf{106.535} \\  \cline{2-9}
    &                       & White-blur\cite{10.1007/978-3-030-66823-5_14}      & \textbf{95.2\%}    & 0.185           & 33.753   & 0.932         & 0.160       & 254.910                                 \\
    &                       & AF\cite{ijcai2022p107}     & 55.5\%      & 0.106     & \underline{37.081}        & \underline{0.969}        & \textbf{0.042}        & \textbf{83.689}\\
    &                       & Saliency-aware\cite{10121622}  & 93.2\%             & 0.113           & 34.589   & 0.961         & 0.090       & 214.399                                 \\
    &                       & Union-aware\cite{10330601}     & 80.7\%             & \underline{0.201}  & 35.443   & 0.947         & 0.122       & 214.399                                 \\
    &                       & DF-RAP\cite{10458678}          & 92.9\%               & \textbf{0.230}          & 34.831              & 0.942              & 0.184           &  133.933                            \\
    \multirow{-15}{*}{AttGAN\cite{8718508}}  & \multirow{-4}{*}{LFW\cite{huang2008labeled}}    & \cellcolor[HTML]{EFEFEF}GRASP & \cellcolor[HTML]{EFEFEF}\underline{95.0\%}         & \cellcolor[HTML]{EFEFEF}0.162          & \cellcolor[HTML]{EFEFEF}\textbf{38.127} & \cellcolor[HTML]{EFEFEF}\textbf{0.975} & \cellcolor[HTML]{EFEFEF}\underline{0.061} & \cellcolor[HTML]{EFEFEF}\underline{103.667} \\ \hline
        &                   & White-blur\cite{10.1007/978-3-030-66823-5_14}      & \textbf{100\%}     & \textbf{0.326}  & 35.491   & 0.958         & 0.046       & 110.017 \\
        &                   & AF\cite{ijcai2022p107}         & \textbf{100\%}      & \underline{0.285}     & 34.937        & 0.951        & 0.050        & 126.872\\
    &                       & Saliency-aware\cite{10121622}  & 98.8\%             & 0.174           & 36.388   & 0.968         & \underline{0.026}       & 91.625                                  \\
    &                       & Union-aware\cite{10330601}     & 94.8\%             & 0.224           & \underline{37.880}   & \underline{0.970}         & 0.035       & \underline{72.600}                                  \\
    &                       & DF-RAP\cite{10458678}          & 54.8\%               & 0.088          & 33.721              & 0.946              & 0.159           &  225.996                            \\
    & \multirow{-5}{*}{CelebA\cite{10.1109/ICCV.2015.425}} & \cellcolor[HTML]{EFEFEF}GRASP & \cellcolor[HTML]{EFEFEF}\textbf{100\%} & \cellcolor[HTML]{EFEFEF}0.244          & \cellcolor[HTML]{EFEFEF}\textbf{41.536} & \cellcolor[HTML]{EFEFEF}\textbf{0.990} & \cellcolor[HTML]{EFEFEF}\textbf{0.006} & \cellcolor[HTML]{EFEFEF}\textbf{31.190} \\ \cline{2-9} 
    &                       & White-blur\cite{10.1007/978-3-030-66823-5_14}      & \textbf{100\%}     & \textbf{0.400}  & 35.969   & 0.952         & 0.108       & 97.071                                 \\
    &                       & AF\cite{ijcai2022p107}         & \textbf{100\%}      & \underline{0.354}     & 37.163        & 0.963        & 0.067        & 66.971\\
    &                       & Saliency-aware\cite{10121622}  & \textbf{100\%}     & 0.241           & 35.005   & 0.960         & 0.027       & 124.514                                 \\
    &                       & Union-aware\cite{10330601}     & 93.0\%             & 0.247           & \underline{37.763}   & \underline{0.972}         & \underline{0.021}       & \underline{75.679}                                  \\
    &                       & DF-RAP\cite{10458678}          & 65.6\%               & 0.085          & 33.632              & 0.954              & 0.119           &  228.563                            \\
    & \multirow{-5}{*}{FFHQ\cite{karras2019style}}   & \cellcolor[HTML]{EFEFEF}GRASP & \cellcolor[HTML]{EFEFEF}\textbf{100\%} & \cellcolor[HTML]{EFEFEF}0.269          & \cellcolor[HTML]{EFEFEF}\textbf{40.854} & \cellcolor[HTML]{EFEFEF}\textbf{0.990} & \cellcolor[HTML]{EFEFEF}\textbf{0.004} & \cellcolor[HTML]{EFEFEF}\textbf{37.618} \\  \cline{2-9} 
    &                       & White-blur\cite{10.1007/978-3-030-66823-5_14}       & \textbf{100\%}     & \textbf{0.338}  & 35.436   & 0.953         & 0.045       & 111.512                                 \\
    &                       & AF\cite{ijcai2022p107}         & 99.2\%      & 0.279     & 35.300        & 0.955        & 0.053        & 103.043\\
    &                       & Saliency-aware\cite{10121622}  & \textbf{100\%}     & 0.219           & 36.698   & 0.972         & 0.021       & 84.874                                  \\
    &                       & Union-aware\cite{10330601}     & 95.3\%             & 0.280           & \underline{38.413}   & \underline{0.971}         & \underline{0.027}       & \underline{65.749}                                  \\
    &                       & DF-RAP\cite{10458678}          & 69.2\%               & 0.096          & 33.760              & 0.944              & 0.155           &  226.924                            \\
    \multirow{-15}{*}{HiSD\cite{9577868}}    & \multirow{-4}{*}{LFW\cite{huang2008labeled}}    & \cellcolor[HTML]{EFEFEF}GRASP & \cellcolor[HTML]{EFEFEF}\textbf{100\%} & \cellcolor[HTML]{EFEFEF}\underline{0.319}          & \cellcolor[HTML]{EFEFEF}\textbf{41.442} & \cellcolor[HTML]{EFEFEF}\textbf{0.989} & \cellcolor[HTML]{EFEFEF}\textbf{0.006} & \cellcolor[HTML]{EFEFEF}\textbf{33.303} \\ 
    \hline
    \end{tabular}
\end{table*}

\subsubsection{Hyperparameter Settings}

All the images in the experiments are scaled to a resolution of $256\times 256$ pixels. 
The kernel size used for Gaussian smoothing is set to 11. 
The perturbation range $\epsilon$ is set to 0.05. 
In Eqs.~\eqref{eq:gs1}-\eqref{eq:gs3}, $\lambda_1$, $\lambda_2$, and $\lambda_3$ are set to 10, 5, and 1, respectively; and $\mu_1$, $\mu_2$, and $\mu_3$ are all set to 1.
In Eq.~\eqref{eq:G_total}, $\eta_1$, $\eta_2$, and $\eta_3$ are set to 11, 3, and 19, respectively. 
In Eq.~\eqref{eq:update_adv_image}, $\kappa$ is set to 10. The number of iterations $T$ is set to 20. 

\subsection{Performance Analysis}
\label{Performance Comparison}

\subsubsection{Effectiveness across Models, Datasets and Scales}
To evaluate the effectiveness of the proposed method under varying settings, we perform experiments using different deepfake models and image scales, where 50, 100, 500, and 1000 images are randomly selected for testing. As shown in Fig.~\ref{three_images}, GRASP maintains consistently high performance across all settings. For both StarGAN and HiSD, the DSR remains at 100\%, with PSNR around 40\,dB and SSIM close to 0.99. The results demonstrate that GRASP is consistently effective across different models and data scales. Notably, performance with 100 images is relatively better balanced across metrics, so subsequent evaluations are based on this setting.

\subsubsection{Defense Effectiveness (RQ1)}
The effectiveness of GRASP in disrupting deepfake manipulations is evidenced by a comprehensive comparison with SOTA methods across multiple deepfake models and datasets, as summarized in Table~\ref{table1}. 
The results in Table~\ref{table1} demonstrate that, compared to SOTA methods, GRASP consistently achieves superior or highly competitive DSRs across all datasets and models. Notably, it outperforms Union-aware~\cite{10330601} and DF-RAP~\cite{10458678} in nearly all settings. Specifically, GRASP surpasses Union-aware by at least 14.3\% on the AttGAN model and achieves a DSR nearly six times higher than AF. On the HiSD model, GRASP outperforms DF-RAP by no less than 30.8\%. Although GRASP does not yield the highest $L_2$ distances among the compared methods, it maintains competitive values. 
This strong cross-model performance highlights the generalizability of GRASP as a proactive defense method. 



\begin{figure*}[!t]
    \centering
    \includegraphics[width=\textwidth]{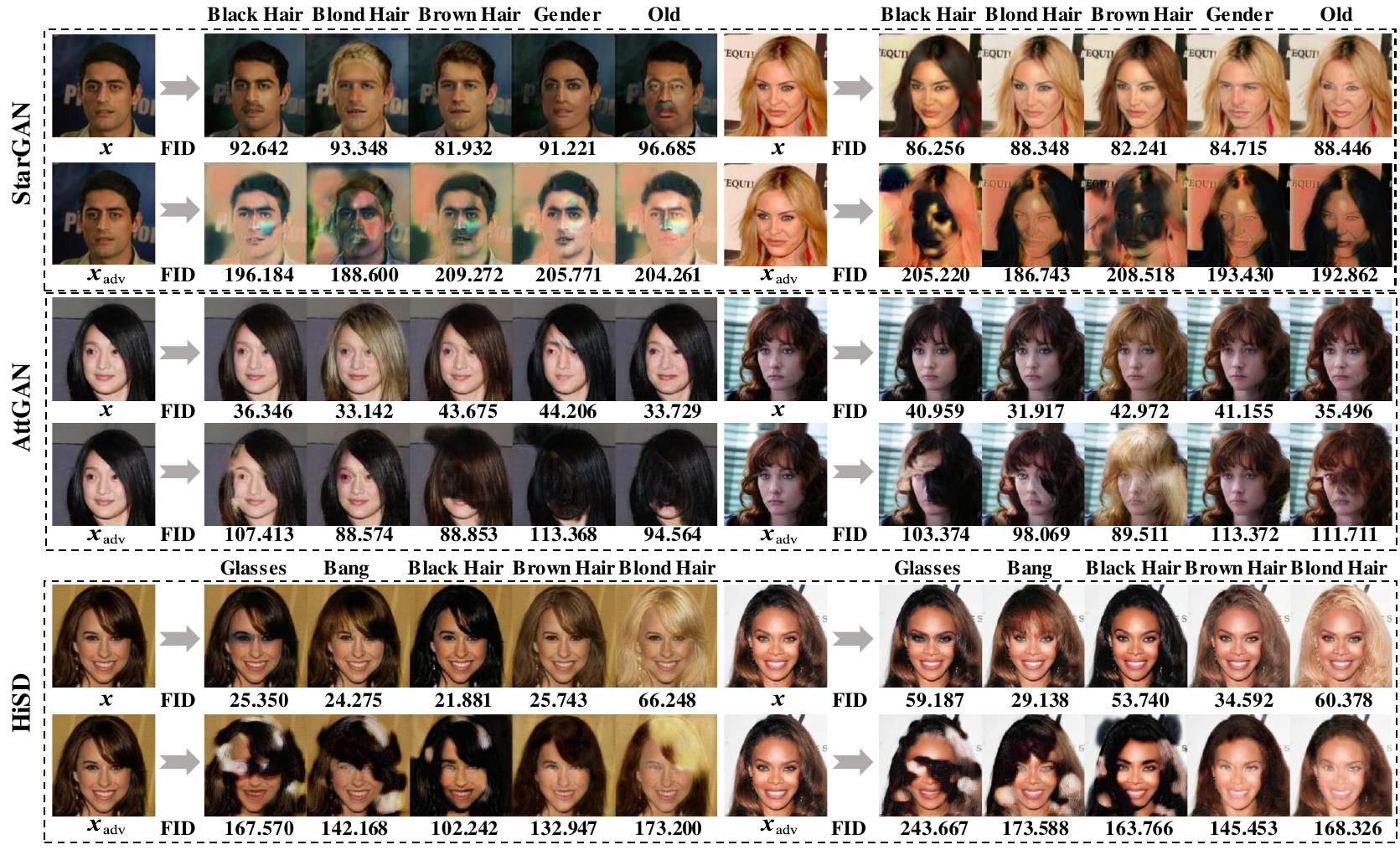}
    \captionsetup{justification=centering}
    \caption{Visualization examples of disrupting attribute editing. For each target model, the first row shows the deepfake model's forgery results on the original images, while the second row displays the deepfake model's output on the adversarial facial images. }
    \label{fig5}
\end{figure*}

Visual comparisons in Fig.~\ref{fig5} further illustrate the impact of the generated perturbations, highlighting how adversarial images differ from their original counterparts when processed by various deepfake models. 
The adversarial images visibly distort the outputs of deepfake models, resulting in attribute editing outputs that appear unrealistic or semantically inconsistent. This is accompanied by a noticeable increase in FID scores, indicating that GRASP effectively disrupts the manipulation process and prevents deepfake models from producing convincing synthetic outputs.

\begin{figure}[!t]
    \centering
    \includegraphics[width=\linewidth]{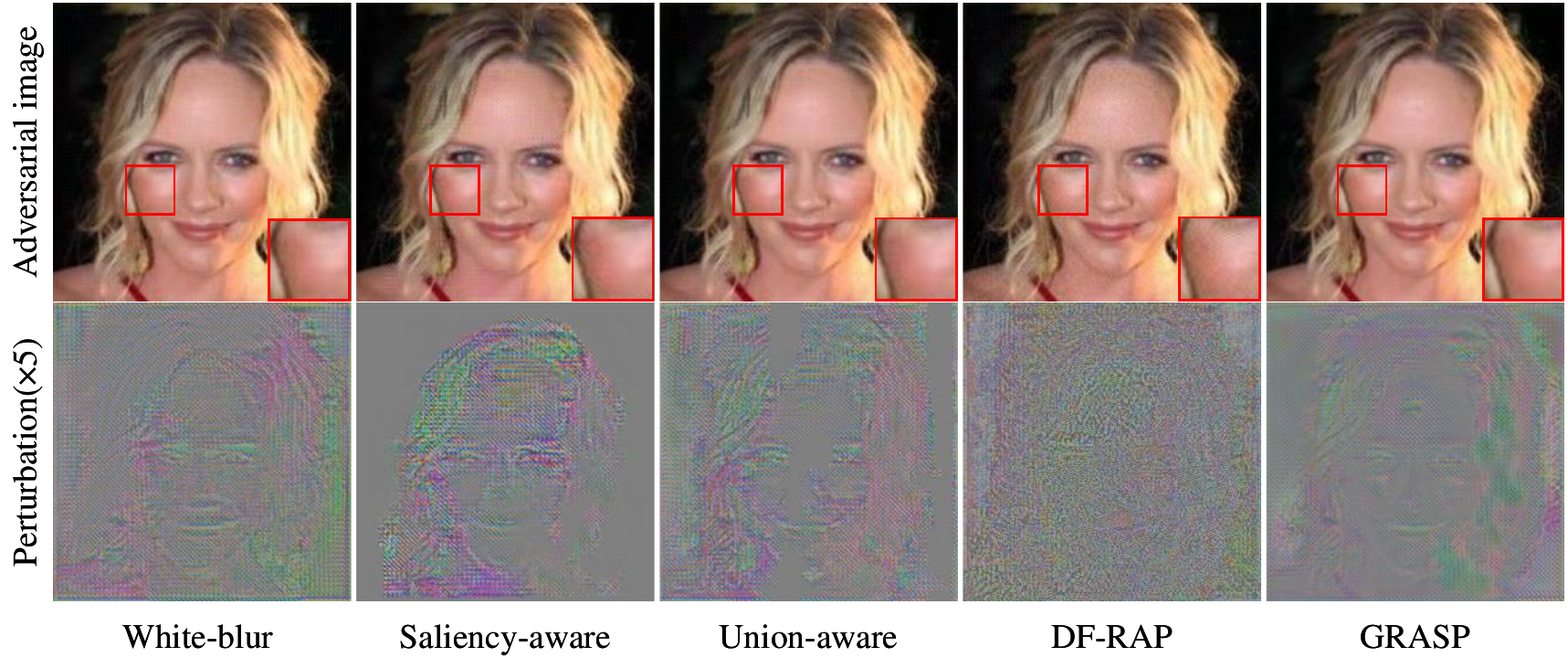}
    \caption{Comparison of the visualized  perturbations generated by GRASP and the methods in \cite{10.1007/978-3-030-66823-5_14}, \cite{10121622}, \cite{10330601} and\cite{10458678}. For each adversarial facial image produced by different models, detailed regions are magnified and highlighted within the red box. }
    \label{fig6}
\end{figure}

\begin{table*}[htbp]
    \centering
    \caption{Robustness evaluation of GRASP and SOTA methods against post-processing operations.}
    \label{table2}
    \renewcommand\arraystretch{1.2} 
    \begin{tabular}{llcccc|cccc|cccc}
    \hline
 &                    & \multicolumn{4}{c}{Gaussian Blur (kernel size)} & \multicolumn{4}{c}{Average Blur (kernel size)} & \multicolumn{4}{c}{Rotation (angle)} \\ \cline{3-14} 
    \multirow{-2}{*}{Methods}   & \multirow{-2}{*}{} & 1          & 3         & 5         & 7         & 1         & 3         & 5         & 7         & 45      & 90      & 135    & 180    \\ \hline
    \multirow{2}{*}{White-blur\cite{10.1007/978-3-030-66823-5_14}}  & DSR↑   & 100\%      & 100\%     & 100\%     & 100\%     & 100\%     & 100\%     & \textbf{57.4\%}    & \textbf{49.8\%}    & 100\%   & 100\%   & 100\%  & 100\%  \\
         & $L_2$↑                & 0.578      & 0.509     & 0.502     & 0.494     & 0.577     & 0.310     & 0.063     & 0.058     & 0.322   & 0.484   & 0.419  & 0.511  \\ \hline
     \multirow{2}{*}{AF\cite{ijcai2022p107}}  & DSR↑  & 100\%    & 3.2\%   & 0.0\%   & 0.0\%   & 100\%   & 17.2\%  & 0.4\%    & 1.6\%    & 100\%   & 100\%   & 100\%  & 100\%  \\
                                                    & $L_2$↑                & 0.968      & 0.013     & 0.005     & 0.001     & 0.968     & 0.035     & 0.011     & 0.017     
                                                    & 0.342   & 0.342   & 0.415  & 0.459  \\ \hline
    \multirow{2}{*}{Saliency-aware\cite{10121622}}                                                & DSR↑               & 100\%      & 100\%     & 100\%     & 100\%     & 100\%     & 99.2\%    & 26.8\%    & 20.8\%    & 100\%   & 100\%   & 100\%  & 100\%  \\
                                                    & $L_2$↑                & 0.357      & 0.292     & 0.279     & 0.276     & 0.357     & 0.204     & 0.043     & 0.042     & 0.332   & 0.434   & 0.418  & 0.472  \\ \hline
    \multirow{2}{*}{Union-aware\cite{10330601}}                                                & DSR↑               & 98.8\%     & 86.9\%    & 89.0\%    & 90.2\%    & 98.8\%    & 77.6\%    & 26.5\%    & 25.7\%    & 100\%   & 100\%   & 100\%  & 100\%  \\
                                                    & $L_2$↑                & 0.371      & 0.289     & 0.314     & 0.330     & 0.431     & 0.204     & 0.044     & 0.041     & 0.337   & 0.434   & 0.405  & 0.483  \\ \hline
    \multirow{2}{*}{DF-RAP\cite{10458678} }                                                & DSR↑               & 100\%      & 3.6\%     & 3.2\%     & 3.2\%     & 100\%     & 2.8\%     & 1.2\%     & 2.0\%     & 100\%   & 100\%   & 100\%  & 100\%  \\
                                                    & $L_2$↑                & 0.274      & 0.013     & 0.014     & 0.013     & 0.274     & 0.014     & 0.011     & 0.019     & 0.326   & 0.399   & 0.411  & 0.502  \\ \hline
    \rowcolor[HTML]{EFEFEF} 
                            & DSR↑               & \textbf{100\%}      & \textbf{100\%}     & \textbf{100\%}     & \textbf{100\%}     & \textbf{100\%}     & \textbf{100\%}     & 26.0\%    & 27.6\%    & \textbf{100\%}   & \textbf{100\%}   & \textbf{100\%}  & \textbf{100\%}  \\
    \rowcolor[HTML]{EFEFEF} 
    \multirow{-2}{*}{\cellcolor[HTML]{EFEFEF}GRASP} & $L_2$↑                & 0.470      & 0.363     & 0.303     & 0.298     & 0.470     & 0.248     & 0.043     & 0.042     & 0.329   & 0.447   & 0.423  & 0.447  \\ \hline
    \end{tabular}
\end{table*}

\subsubsection{Perturbation Imperceptibility (RQ2)}
Visual imperceptibility of the adversarial perturbations is evaluated using several standard metrics, including PSNR, SSIM, LPIPS and LF, as reported in Table~\ref{table1}. GRASP achieves superior performance across nearly all visual metrics on both AttGAN and HiSD models. For instance, on HiSD-CelebA, GRASP attains a PSNR of 41.536\,dB, an SSIM of 0.99, an LPIPS of 0.004, and an LF value of 37.618, outperforming all compared methods. These results confirm that GRASP introduces minimally perceptible perturbations while maintaining strong defense effectiveness. Although AF achieves slightly better visual quality than GRASP on the StarGAN model, this advantage is not observed across other architectures. In contrast, GRASP consistently maintains a favorable balance between visual fidelity and defense performance across diverse generative models, demonstrating better generalizability.

Beyond quantitative metrics, Fig.~\ref{fig6} presents visual comparisons of the residual perturbation between the adversarial and original images. Compared to White-blur\cite{10.1007/978-3-030-66823-5_14}, Saliency-aware\cite{10121622}, Union-aware\cite{10330601} and DF-RAP\cite{10458678}, the perturbations generated by GRASP are more uniformly distributed and visually smoother, without introducing noticeable artifacts. This visual subtlety confirms the high imperceptibility achieved by GRASP.

\subsubsection{Robustness}
The robustness of GRASP against common post-processing operations is demonstrated through a comparative evaluation with SOTA methods, with results summarized in Table~\ref{table2}. Three representative transformations—Gaussian blur, average blur, and rotation—are selected for this analysis, given their  prevalence in real-world image transmission scenarios. To comprehensively evaluate robustness, we vary the strength of each transformation through adjusting the scale. Specifically, all methods are tested under Gaussian blur and average blur with kernel sizes of 1, 3, 5, and 7, as well as rotation transformations at angles of 45°, 90°, 135°, and 180°. 

As shown in Table~\ref{table2}, GRASP consistently achieves higher DSR and $L_2$ values than most of the SOTA methods, owing to the incorporation of Gaussian smoothing as a noise layer during adversarial image generation. Even when compared to White-blur~\cite{10.1007/978-3-030-66823-5_14}, the most robust baseline, GRASP demonstrates comparable robustness while significantly outperforming it in visual quality, as reflected by visual metrics reported in Table~\ref{table1}. 
It is worth noting, however, that White-blur exhibits stronger robustness under average blur attacks with larger kernel sizes (e.g., 5 and 7). This advantage can be attributed to White-blur’s strategy of embedding most adversarial perturbations in the low-frequency domain, which is less affected by average blur—an attack that primarily suppresses high-frequency components.




\begin{table*}[ht]
    \centering
    \renewcommand{\arraystretch}{1.1}
    \caption{The effectiveness of GRASP against the SimSwap was evaluated on the CelebA dataset.}
    \label{table3}
    \small
    \begin{tabular}{ lccccccc }
    \hline
     Method                                       & DSR↑       & $L_1$↑       & ID sim.↓      & PSNR↑     & SSIM↑     & LPIPS↓     & LF↓  \\
    \hline
    White-blur\cite{10.1007/978-3-030-66823-5_14} & \underline{90\%}        & \underline{0.093}     & \underline{0.263}         & 35.024    & 0.955     & 0.076      & 164.719 \\
    AF\cite{ijcai2022p107} & 26\%        & 0.058     & 0.472         & \underline{39.286}    & \underline{0.980}     & \underline{0.019}      & \textbf{39.179} \\
    Saliency-aware\cite{10121622}                 & \underline{90\%}        & 0.083     & 0.266         & 35.733    & 0.961     & 0.062      & 105.822 \\
    Union-aware\cite{10330601}                    & 62\%        & 0.074     & 0.374         & 35.388    & 0.956     & 0.072      & 134.369 \\
    DF-RAP\cite{10458678}                         & 82\%        & 0.092     & 0.283         & 35.520    & 0.954     & 0.110      & 93.552 \\
    \rowcolor[HTML]{EFEFEF}GRASP               & \textbf{96\%}        & \textbf{0.093}     & \textbf{0.259}         & \textbf{39.559}    & \textbf{0.988}     & \textbf{0.016}      & \underline{52.430} \\
    \hline
    \end{tabular}
\end{table*}

\subsubsection{Effectiveness on Face-Swapping Models}
Face-swapping represents a practically significant form of manipulation,  as it alters identity information rather than semantic attributes. However, this setting remains underexplored in many existing methods, including White-blur~\cite{10.1007/978-3-030-66823-5_14}, Saliency-aware~\cite{10121622}, Union-aware~\cite{10330601} and AF~\cite{ijcai2022p107}. To address this gap and further demonstrate the effectiveness and generalizability of GRASP across diverse deepfake paradigms, we extend our evaluation to the face-swapping model SimSwap~\cite{10.1145/3394171.3413630} and reproduce several SOTA methods for fair and systematic comparison.

In the face-swapping setting, the $L_1$ distance and identity similarity (ID sim.) are employed to evaluate defense effectiveness. Given that face-swapping typically introduces sparse and localized changes in the facial region, the $L_1$ distance is well-suited to capture these sparse pixel-level differences that correspond to meaningful identity changes. Additionally, ID sim. measures the similarity between the original and swapped face images, computed as the cosine similarity between identity embeddings extracted by a face recognition model (e.g., ArcFace~\cite{8953658}). Since face-swapping directly targets identity manipulation, 
a lower ID sim. indicates more effective disruption of identity preservation by the defense. Evaluating the perceptual imperceptibility of the adversarial images, we consistently use PSNR, SSIM, LPIPS, and LF as the evaluation metrics.

\begin{figure}[!t]
\centering
\includegraphics[width=\linewidth]{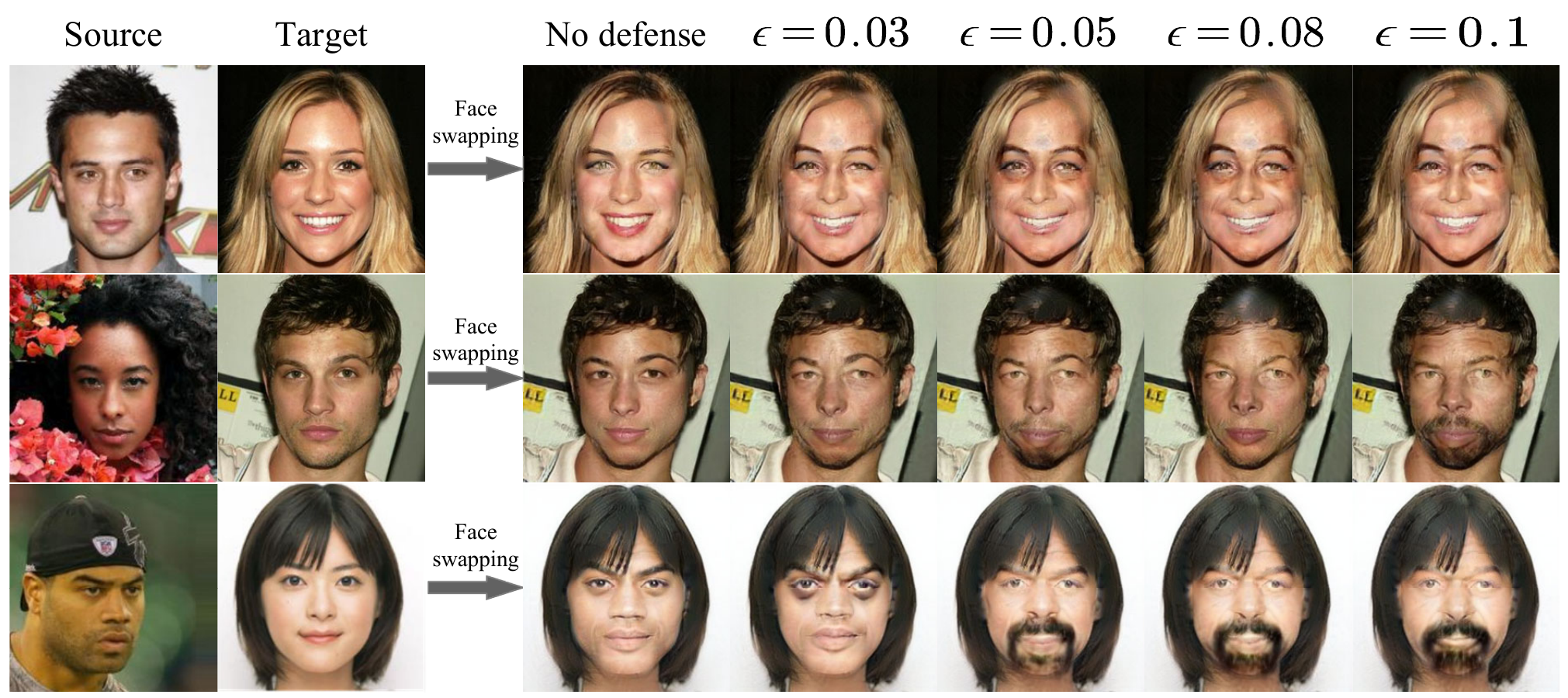}
\caption{The defensive performance of GRASP on the SimSwap model under different $\epsilon$ values. }
\label{fig9}
\end{figure}

Table~\ref{table3} presents the evaluation results of GRASP and four SOTA methods against the face-swapping model SimSwap on the CelebA dataset. As shown in the table, GRASP achieves the highest DSR of 96\%, outperforming AF~\cite{ijcai2022p107} by a substantial margin of 70\%. This significant gap indicates AF is ineffective against face-swapping-based manipulations. Moreover, GRASP attains the lowest ID sim. score of 0.259, indicating effective identity obfuscation. Besides, Fig.~\ref{fig9} provides a visual demonstration of GRASP's defense effectiveness against SimSwap. As the perturbation constraint $\epsilon$ increases, the disruption to identity preservation becomes more pronounced. In terms of perturbation imperceptibility, GRASP consistently ranks first across all perceptual metrics, surpassing other methods by nearly 4dB in PSNR. These results demonstrate that GRASP not only offers the most effective defense against face-swapping but also delivers superior perceptual quality.

In addition, the robustness of the adversarial images generated by each method is evaluated under image distortions in the face-swapping setting. Specifically, we assess performance under Gaussian blur and average blur with increasing kernel sizes, and report the corresponding DSR, $L_1$ distance and ID similarity in Table~\ref{table4}. 
Since AF~\cite{ijcai2022p107} has already demonstrated limited  effectiveness against face-swapping models, its robustness rarely manifests under such attack scenarios.
It is also observed that White-blur~\cite{10.1007/978-3-030-66823-5_14}, Saliency-aware~\cite{10121622}, and Union-aware~\cite{10330601} exhibit an increasing trend in DSR as the Gaussian kernel size grows. This phenomenon can be attributed to the  operational characteristics of the SimSwap model, which primarily modifies low-frequency components of facial images—such as global structure and facial contours—during face synthesis. Since these defense methods also concentrate their perturbations in the low-frequency domain, they are particularly well-suited to disrupt SimSwap’s identity transfer mechanism. As the Gaussian kernel size increases, high-frequency image details are progressively smoothed out, thereby amplifying the relative impact of low-frequency perturbations. This enhances the effectiveness of the adversarial signal embedded by these methods, leading to improved defensive performance. 
Although these methods are well-suited for defending against SimSwap due to their low-frequency perturbation strategies, GRASP still demonstrates stable and competitive defense effectiveness across varying levels of distortion—ranking first in half of the test cases and second only to White-blur in the remaining ones. 


\begin{table}[]
    \centering
    \renewcommand{\arraystretch}{1.1}
    \caption{Robustness of adversarial images under image distortions in the face-swapping setting. }
    \label{table4}
    \resizebox{\linewidth}{!}{
   \begin{tabular}{llcccccccc}
        \hline
                                                        & \multicolumn{1}{c}{}                   & \multicolumn{4}{c}{Gaussian Blur(kernel size)} & \multicolumn{4}{c}{Average Blur(kernel size)} \\ \cline{3-10} 
        \multirow{-2}{*}{Methods}                       & \multicolumn{1}{c}{\multirow{-2}{*}{}} & 1          & 3         & 5         & 7         & 1         & 3         & 5         & 7         \\ \hline
                                                        & DSR↑                                   & 90\%       & 90\%      & \textbf{96\%}      & \textbf{96\%}      & 90\%      & \textbf{96\%}      & \textbf{82\%}      & \textbf{56\%}      \\
                                                        & $L_1$↑                                 & 0.093      & 0.097     & 0.096     & 0.096     & 0.093     & 0.097     & 0.089     & 0.072     \\
        \multirow{-3}{*}{White-blur\cite{10.1007/978-3-030-66823-5_14}}& Id sim.↓                & 0.263      & 0.235     & 0.240     & 0.241     & 0.263     & 0.234     & 0.285     & 0.357     \\ \hline
                                                        & DSR↑                                   & 26\%       & 2\%      & 2\%      & 2\%      & 26\%      & 0\%      & 2\%      & 4\%      \\
                                                        & $L_1$↑                                 & 0.057      & 0.025     & 0.025     & 0.025     & 0.0569     & 0.021     & 0.025     & 0.032     \\
        \multirow{-3}{*}{AF\cite{ijcai2022p107}}& Id sim.↓                & 0.469      & 0.666     & 0.665     & 0.665     & 0.469     & 0.681     & 0.654     & 0.597     \\ \hline
                                                        & DSR↑                                   & 92\%       & 94\%      & 94\%      & 94\%      & 92\%      & 94\%      & 60\%      & 42\%      \\
                                                        & $L_1$↑                                 & 0.083      & 0.086     & 0.084     & 0.084     & 0.083     & 0.086     & 0.072     & 0.060     \\
        \multirow{-3}{*}{Saliency-awre\cite{10121622}}                 & Id sim.↓                & 0.281      & 0.259     & 0.275     & 0.276     & 0.281     & 0.260     & 0.352     & 0.403     \\ \hline
                                                        & DSR↑                                   & 62\%       & 62\%      & 66\%      & 68\%      & 62\%      & 68\%      & 54\%      & 34\%      \\
                                                        & $L_1$↑                                 & 0.074      & 0.078     & 0.078     & 0.078     & 0.074     & 0.078     & 0.071     & 0.059     \\
        \multirow{-3}{*}{Union-aware\cite{10330601}}                   & Id sim.↓                & 0.374      & 0.342     & 0.345     & 0.345     & 0.374     & 0.342     & 0.386     & 0.438     \\ \hline
                                                        & DSR↑                                   & 82\%       & 22\%      & 18\%      & 18\%      & 82\%      & 14\%      & 10\%      & 10\%      \\
                                                        & $L_1$↑                                 & 0.092      & 0.066     & 0.062     & 0.062     & 0.092     & 0.057     & 0.048     & 0.043     \\
        \multirow{-3}{*}{DF-RAP\cite{10458678}}                        & Id sim.↓                & 0.283      & 0.474     & 0.493     & 0.495     & 0.283     & 0.522     & 0.554     & 0.534     \\ \hline
        \rowcolor[HTML]{EFEFEF} 
        \cellcolor[HTML]{EFEFEF}                        & DSR↑                                   & \textbf{96\%}       & \textbf{96\%}      & \underline{94\%}      & \underline{94\%}      & \textbf{96\%}      & \underline{94\%}      & \underline{64\%}      & \underline{50\%}      \\
        \rowcolor[HTML]{EFEFEF} 
        \cellcolor[HTML]{EFEFEF}                        & $L_1$↑                                 & 0.093      & 0.096     & 0.095     & 0.095     & 0.093     & 0.095     & 0.083     & 0.064     \\
        \rowcolor[HTML]{EFEFEF} 
        \multirow{-3}{*}{\cellcolor[HTML]{EFEFEF}GRASP} & Id sim.↓                               & 0.259      & 0.252     & 0.259     & 0.260     & 0.259     & 0.257     & 0.320     & 0.401     \\ \hline
    \end{tabular}
    }
\end{table}

\begin{table*}[htp]
    \centering
    \renewcommand\arraystretch{1.1} 
    \caption{Ablation study results validating the contribution of each individual component in GRASP}
    \label{table5}
    \small
    \begin{tabular}{ lllllll }
    \hline
    Loss      & DSR↑  & $L_2$↑    & PSNR↑  & SSIM↑  & LPIPS↓     & LF↓   \\ \hline
    $L_{\text{MSE}}$    & 100\% & 0.504  & 34.164 & 0.938  & 0.031      & 119.405 \\
    $L_{\text{MSE}}$+$L_{\text{SSIM}}$    & 100\% & 0.455  & 35.846 & 0.963  & 0.018      & 88.676 \\
    $L_{\text{MSE}}$+$L_{\text{SSIM}}$+$L_{\text{LF}}$   & 24.8\% & 0.040 & 44.237 & 0.994  & 0.002      & 5.287 \\
    $L_{\text{MSE}}$+$L_{\text{SSIM}}$+$L_{\text{LF}}$+Gradient Projection & 100\% & 0.465 & 39.783 & 0.986  & 0.008      & 42.452 \\
    \hline
    \end{tabular}
\end{table*}

\subsection{Ablation Study}
\label{Ablation Study}

An ablation study is presented to clarify the role of each loss component in Eq.~\eqref{eq:total_loss} and the gradient projection introduced in Section~\ref{Gradient Projection} for balancing robustness and perceptual quality within the GRASP framework. As shown in Table~\ref{table5}, different combinations of \(L_{\text{MSE}}\), \(L_{\text{SSIM}}\), \(L_{\text{LF}}\) and the gradient projection module are evaluated in terms of DSR, PSNR, SSIM, LPIPS, $L_2$ and LF metrics. Based on the results in the table, the following observations can be made.



When using only \(L_{\text{MSE}}\) as the supervision signal—which is also the sole loss function employed in White-blur~\cite{10.1007/978-3-030-66823-5_14}— the generated adversarial images achieve high DSR, but suffer from limited perceptual quality.
Incorporating the SSIM loss term, resulting in the combined loss \(L_{\text{MSE}}\) + \(L_{\text{SSIM}}\), aligns with the mainstream loss configuration adopted in many proactive defense methods. 
This combination leads to modest improvements across all visual quality metrics.
As elaborated in Eq.~\eqref{eq:total_loss}, GRASP introduces the LF loss term to enhance perceptual imperceptibility. The strong visual quality achieved—such as a 29.5\% boost in PSNR, as reported in Table~\ref{table5}—demonstrates the effectiveness of incorporating $L_{\text{lf}}$. Whereas, without consoling the underlying gradient conflict among loss terms, the DSR drops significantly to 24.8\%, highlighting the necessity of conflict-aware optimization.   
Hence, by introducing the gradient projection strategy elaborated in Section \ref{Gradient Projection}, both perceptual quality and defense effectiveness are simultaneously improved, with the DSR regaining at 100\%. 

\begin{figure}
\centering
\includegraphics[width=\linewidth]{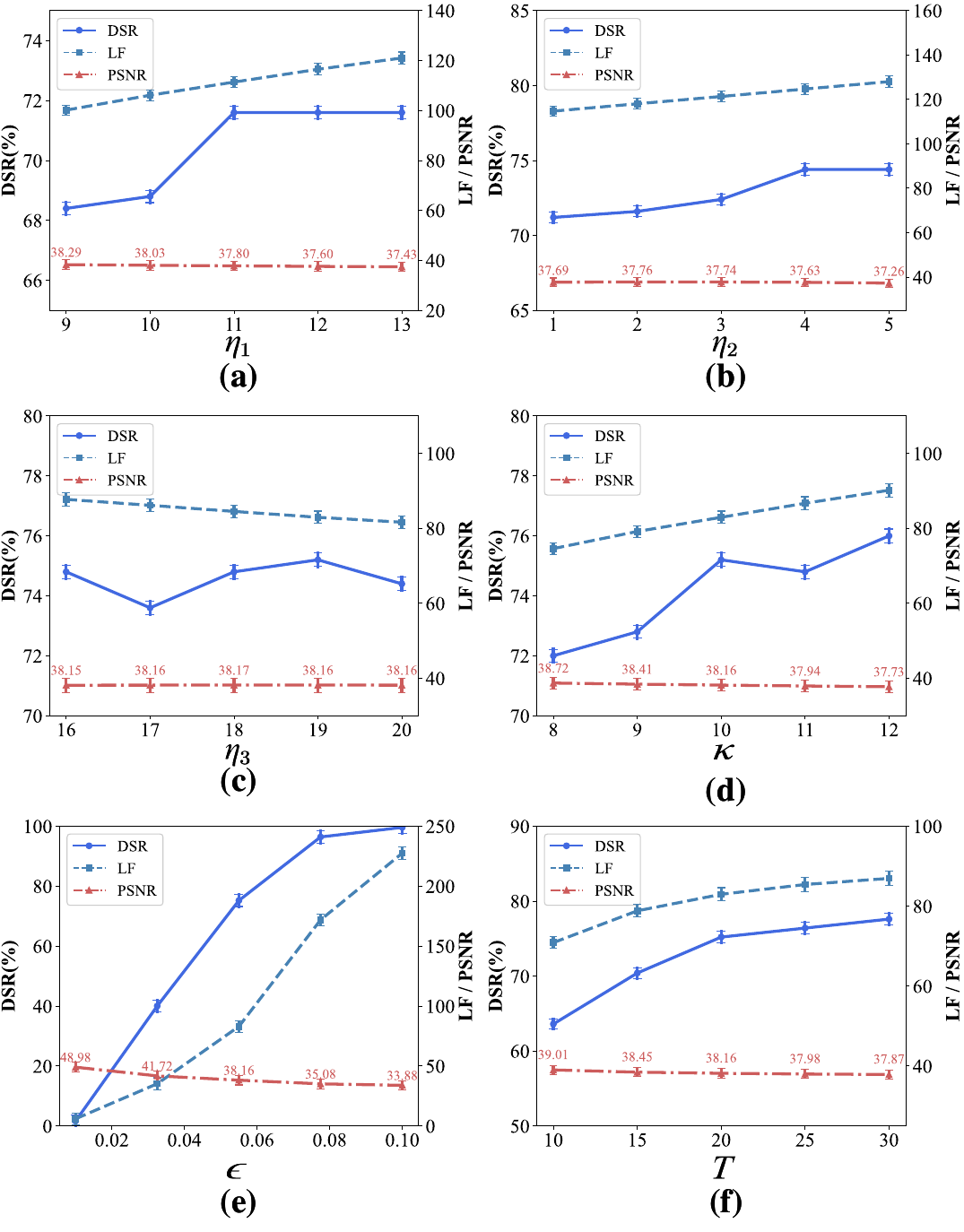}
\caption{
DSR, PSNR, and LF results for adversarial facial images generated under varying values of each individual hyperparameter. A total of six hyperparameter are considered: gradient weighting coefficients ($\eta_1$, $\eta_2$, and $\eta_3$), image processing parameters ($\kappa$), perturbation strength ($\epsilon$), and the number of optimization iterations ($T$).
}
\label{fig8}
\end{figure}

\subsection{Rationale of Hyperparameters Setting}
To validate the rationale behind the key hyperparameter settings in the proposed method and to further analyze their impact on defense performance and image quality, we conducted a series of adversarial image generation experiments under various configurations. Six representative hyperparameter groups were selected, covering gradient weighting coefficients ($\eta_1$, $\eta_2$, and $\eta_3$), image processing parameters ($\kappa$), perturbation strength ($\epsilon$), and the number of optimization iterations ($T$). The experimental results are summarized in Fig.~\ref{fig8}, and the findings are discussed as follows.


The gradient weighting coefficients $\eta_1$, $\eta_2$, and $\eta_3$ were analyzed to understand their influence on optimization behavior. As shown in Fig.~\ref{fig8}(a), increasing $\eta_1$ from 9 to 11 leads to a significant improvement in DSR, which then plateaus for values beyond 11. However, this increase also causes a rise in LF energy and a corresponding drop in PSNR, indicating reduced perceptual quality. Similarly, $\eta_2$, varied between 1 and 5 (Fig.~\ref{fig8}(b)), shows a consistent increase in both LF and DSR, while PSNR decreases slightly. To prevent over-enhancement of perturbation magnitude, we select $\eta_2=3$ for a balanced trade-off. For $\eta_3$, explored within the range of 16–20 (Fig.~\ref{fig8}(c)), the best DSR is achieved at $\eta_3=19$, while PSNR improves steadily and LF decreases as the value increases. Based on these observations, we adopt $\eta_1=11$, $\eta_2=3$, and $\eta_3=19$ as the optimal configuration to ensure effective defense with minimal degradation in visual quality.




We further investigate the impact of image processing configuration $\kappa$, perturbation strength $\epsilon$, and the number of optimization iterations $T$ on overall performance. As shown in Fig.~\ref{fig8}(d), increasing $\kappa$ from 8 to 12 improves the DSR, reaching its peak at $\kappa=12$, but at the cost of higher LF values and significantly reduced PSNR. To balance robustness and visual quality, $\kappa=10$ is selected. For perturbation strength $\epsilon$, tested in the range $[0, 0.1]$ (Fig.~\ref{fig8}(e)), DSR improves rapidly when $\epsilon \leq 0.05$, while LF increases sharply and PSNR drops when $\epsilon$ exceeds 0.05. Thus, we set $\epsilon=0.05$ to ensure effective defense while maintaining imperceptibility. As for the number of optimization iterations $T$, results in Fig.~\ref{fig8}(f) show that while DSR saturates beyond $T=20$, both LF and PSNR continue to degrade, indicating worsening visual quality. Therefore, $T=20$ is chosen as the optimal setting to achieve a good trade-off between computational efficiency, visual fidelity, and defense effectiveness.

\section{Conclusion}
\label{conclusion}
In this work, we propose GRASP, a gradient-projection-based adversarial defense method designed to disrupt deepfake manipulations while maintaining high visual fidelity. Unlike existing methods that often trade off imperceptibility for robustness, GRASP achieves a fine-grained balance by integrating structural similarity and low-frequency perceptual constraints into the optimization process. To resolve gradient conflicts arising from multi-objective supervision, we introduce a novel cross-gradient projection strategy, enabling stable convergence and effective defense across multiple deepfake paradigms.
Experimental results demonstrate that GRASP achieves strong defense success rates and superior visual quality compared to state-of-the-art methods. Ablation studies further validate the contributions of each component and confirm the method’s effectiveness across key hyperparameter settings.
In future research, we plan to extend GRASP to video-based deepfake scenarios, where maintaining temporal consistency is crucial. We also aim to evaluate its robustness against emerging generative architectures to ensure long-term applicability.

\clearpage
\bibliographystyle{IEEEtran}
\bibliography{references.bib}

\end{document}